\documentclass[twocolumn,showpacs,preprintnumbers,amsmath,amssymb]{revtex4}

\usepackage{graphicx}
\usepackage{dcolumn}
\usepackage{bm}

\begin{document}

\title{Neutrino neutral reaction on $^{4}{\rm He}$, effects of final
state interaction and realistic NN force}

\author{Doron Gazit}
\affiliation{The Racah Institute of Physics, The Hebrew University, 91904
Jerusalem, Israel.}
\email{gdoron@phys.huji.ac.il}
\author{Nir Barnea}
\affiliation{The Racah Institute of Physics, The Hebrew University, 91904
Jerusalem, Israel.}

\date{\today}

\begin{abstract}
The inelastic neutral reaction of neutrino on $^{4}\mathrm{He}$ is
calculated microscopically, including full final state interaction among
the four nucleons. The calculation is performed using the Lorentz integral
transform (LIT) method and the hyperspherical-harmonic effective interaction
approach (EIHH), with a realistic nucleon-nucleon interaction. A
detailed energy dependent calculation is given in the impulse
approximation. With respect to previous calculations, this work
predicts an increased reaction cross-section by $10\%-30\%$ for
neutrino temperature up to $15$ MeV. 
\end{abstract}

\maketitle
The interest in neutrino reactions with nuclear targets stems from
the role they play in major questions of contemporary physics. Such
reactions are of central importance in various astrophysical phenomena, such
as supernova explosion and the nucleosynthesis of the elements. In this
letter, we present a microscopic ab-initio calculation of the neutral
inelastic reactions of $^{4}\mathrm{He}$ with $\nu _{x}(\overline{\nu _{x}})$ 
($x=e,\mu ,\tau $).

Core collapse supernovae are widely accepted to be a neutrino driven
explosion of a massive star. When the iron core of a massive star becomes
gravitationally unstable it collapses until short-range nuclear forces halt
the collapse and drive an outgoing shock through the outer layers of the
core and the inner envelope. However, the shock loses energy through
dissociation of iron nuclei and neutrino radiation, and gradually stalls, it
becomes an accretion shock. It is believed, but to date not proven, that the
shock is then revived as neutrinos emitted from the collapsed core (the
proto-neutron star) deposit energy in the collapsing layers to overcome the
infall and eventually reverse the flow to an outgoing shock which explodes
the star. Hydrodynamic simulations of a collapsing star, which are
restricted to spherical symmetry, fail in reviving the shock \cite{LI01}.
Lately it was shown \cite{BU03} that even in full 2-D calculations the shock
is not revived. In order to revive the shock, the neutrinos must deposit
about $1\%$ of their energy in the matter behind the shock. The
latter, which is assumed to be in thermodynamic equilibrium, is composed
mainly of protons, neutrons, electrons, and $^{4}\mathrm{He}$ nuclei. In
contrast to the fairly known cross-sections of neutrinos with electrons and
nucleons, the interaction of neutrinos with $^{4}\mathrm{He}$ is not
accurately known, and to date there is no realistic microscopic calculation
of the inelastic $^{4}\mathrm{He}$-neutrino cross-section. The effect of
neutrino-$^{4}\mathrm{He}$ interaction on the delayed shock mechanism was
investigated by Bruenn and Haxton \cite{BR91}, through a presupernova 1-D model
of a $15\;M_\odot$ star. In that model, they found only a small
reheating of the matter behind the shock, which can be attributed to
the low mean energy of the neutrinos in comparison to the high
threshold energy of the Alpha nucleus.
This conclusion may change with different progenitor,
or with enlarged inelastic neutrino-$^{4}\mathrm{He}$ cross-sections.

The neutrinos migrating out of the proto-neutron star are in flavor equilibrium
for most of their migration. The electron-neutrinos remain in equilibrium
with matter for a longer period than their heavy-flavor counterparts, due to
the larger cross sections for scattering of electrons and because of charge
current reactions. Thus the heavy-flavor neutrinos decouple from deeper
within the star, where temperatures are higher. Typical calculations yield
temperatures of $\sim 10$MeV for $\mu $- and $\tau $- neutrinos
\cite{WI88}, which is approximately twice the temperature of
electron-neutrinos.
Consequently, there is a considerable amount of $\nu _{\mu ,\tau }$ with
energies above $20$ MeV that can dissociate the $^{4}\mathrm{He}$ through
neutral reaction. 

The flux of neutrinos emitted in the collapse process is sufficiently
large to initiate nucleosynthesis in the overlaying shells of heavy
elements. Neutral reactions of Alpha and neutrino in the inner Helium
shell are part of reaction sequences leading to the production of the
rare $A=7$ Lithium and Beryllium isotopes \cite{WO90},
\cite{WO95}. Thus, better understanding of the $\nu-\alpha$ reaction
can lead to better prediction for the abundances of these elements. 

Theoretical understanding of neutrino-nucleus scattering process is
achieved through perturbation theory of the weak
interaction model. The nuclear electroweak transition operator
consists of one- and
many-body components. The many-body currents are a result of meson
exchange between the nucleons, and usually contribute up
to $10\%$ of the cross-section, in the supernova energy
regime. However, when leading one-body terms are suppressed their
contribution can be even larger. The
current work is done in the impulse approximation, thus taking into
account only one-body terms. 
The one-body currents connect the $^{4}\mathrm{He}$ ground
state and final state wave functions. In order to calculate the
cross-section in a percentage level accuracy, one needs a solid
estimate of these wave functions. Alas, for nuclear systems with more
than three constituents, where particle correlation plays a decisive
role, the computation of intermediate-energy continuum wave function
is currently out of reach.

To facilitate the calculation of the neutral reaction of neutrino and
alpha particle we introduce several modern methods. The
calculation of the nuclear dynamics is carried out by combining two
powerful tools: the Lorentz integral transform (LIT) method
\cite{EF94} and the effective interaction hyperspherical
harmonics (EIHH) method \cite{BA00}. First we use the LIT
method in order to convert the scattering problem into a bound
state like problem, and then the EIHH method is used
to solve the resulting equations. Using this procedure we solve the
final state interaction problem avoiding continuum wave
functions. This method was used successfully to calculate the
photoabsorption cross sections of up to six body nuclei
\cite{BA02}. To this end we use nuclear Hamiltonian consists
of the realistic nucleon-nucleon potential AV8'.

In the limit of small momentum transfer (compared to the Z particle rest
mass), the effective Hamiltonian can be written as

\begin{equation}
\hat{H}_{W}=\frac{G}{\sqrt{2}}\int {d^{3}xj_{\mu }(\vec{x})J^{\mu }(\vec{x})}
\label{eq:hamil}
\end{equation}

\begin{table}[t] 
\begin{tabular}{cccc}
\hline
\hline
T [MeV] &  \multicolumn{2}{c}{$\langle \sigma \rangle_T$
[$10^{-42}cm^{2}$] }  &  \hspace{0.5cm} $\langle \sigma \omega \rangle_T$ \\
 & \hspace{0.15cm} This work & \hspace{0.25cm} Ref.~\cite{WO90} 
 & \hspace{0.5cm} $[10^{-40}cm^{2}\rm{MeV}] $\\ 
\hline
 4    &  2.09(-3) &    -     & 5.27(-4) \\
 6    &  3.84(-2) & 3.87(-2) & 1.03(-2) \\
 8    &  2.25(-1) & 2.14(-1) & 6.30(-2) \\
 10   &  7.85(-1) & 6.78(-1) & 2.30(-1) \\
 12   &  2.05     & 1.63     & 6.27(-1) \\
 14   &  4.45     &    -     & 1.42     \\
 16   &  8.52     &    -     & 2.84     \\
\hline
\hline
\end{tabular}
\caption{{\label{tab:crs}} Flavor and temperature averaged
inclusive inelastic cross-section and energy transfer cross-section
calculated. The temperatures are given in MeV, the cross-sections in
$10^{-42}cm^{2}$, and the energy transfer cross-sections in
$10^{-40}cm^{2}MeV$}
\end{table}

where $G$ is the Fermi weak coupling constant, $j_{\mu }(\vec{x})$ is the
leptonic current, and $J^{\mu }$ is the hadronic current. The matrix element
of the leptonic current is $\langle f|j_{\mu }|i\rangle =l_{\mu }e^{-i\vec{q}
\cdot \vec{x}}$, where ${l}_{\mu }=\bar{u}(k_{\nu ^{\prime }})\gamma _{\mu
}(1-\gamma _{5})u(k_{\nu })$. The nuclear current, 
\begin{equation}
J_{\mu }^{hadronic}=(1-2\cdot \sin ^{2}\theta _{W})\frac{\tau _{0}}{2}J_{\mu
}+\frac{\tau _{0}}{2}\vec{J}_{\mu }^{5}-2\cdot \sin ^{2}\theta _{W}\frac{1}{2
}J_{\mu },  \label{eq:hadcu}
\end{equation}
consists of one body weak currents, but also many body corrections due to
meson exchange. In this work we use the impulse approximation. Since
the momentum transfer relevant to our calculation are small compared
to the nucleon mass, we ignore relativistic corrections. The
differential cross-section is given by Fermi's golden rule. Thus, in
order to consider recoil effects, and with unoriented and unobserved
targets, the differential cross-section takes the form,
\begin{eqnarray}
\lefteqn{d\sigma = \int d\epsilon \delta (\epsilon -\omega +\frac{q^{2}}
{2M_{^{4}\mathrm{He}}})
2\pi \frac{d^{3}\vec{k}_{f}}{(2\pi )^{3}}}  \\ & & {\sum_{f}\frac{
\sum_{M_{i}=-J_{i}}^{J_{i}}}{2J_{i}+1}
\sum_{helicities}|\langle f|
\hat{H}_{W}|i\rangle |^{2}\delta (E_{f}-E_{i}+\epsilon )} \nonumber
\end{eqnarray}%
where $\vec{k}_{f}$ is the momentum of the outgoing neutrino, $\omega $
is the energy transfer, and $\vec{q}$ is the momentum transfer.

Choosing the $\hat{z}$ direction to be parallel to the momentum transfer,
and $\theta $ to be the angle between the incoming neutrino direction and
outgoing neutrino direction, the cross-section can be written as \cite{DO76},
\begin{widetext}
\begin{eqnarray}
{\frac{d\sigma }{dk_{f}}=\int d\epsilon \delta (\epsilon -\omega +\frac
{q^{2}}{2M_{^{4}\mathrm{He}}})\frac{4G^{2}}{2J_{i}+1}k_{f}^{2}
\int_{0}^{\pi }\sin {\theta }d\theta \left\{ \left[ \sin ^{2}\frac{\theta 
}{2}-\frac{q^{\mu }q_{\mu }}{2q^{2}}\cos ^{2}\frac{\theta }{2}\right]
\sum_{J\geq 1}\left[ {R}_{\hat{M}_{J}}(\epsilon )+{R}_{\hat{E}_{J}}(\epsilon )
\right] \right. }   \nonumber
\label{eq:crs} \\
{\left. \mp {\sin \frac{\theta }{2}}\sqrt{\sin ^{2}\frac{\theta }{2}-\frac{
q^{\mu }q_{\mu }}{2q^{2}}\cos ^{2}\frac{\theta }{2}}\sum_{J\geq 1}2{R}_{\hat{
E}_{J}\hat{M}_{J}}(\epsilon )+\cos ^{2}\frac{\theta }{2}\sum_{J\geq 0}R_{\hat{C
}_{J}-\frac{\omega }{q}{\hat{L}}_{J}}(\epsilon )\right\} }
\end{eqnarray}
\end{widetext}
the $-$ ($+$) is for neutrino (anti-neutrino). The functions 
\begin{eqnarray}
\lefteqn{R_{\hat{O}_{1}\hat{O}_{2}}(\omega )=\int d\Psi _{f}} \\ & & 
{\langle \Psi_{0}\mid \mid \hat{O}_{1}\mid \mid \Psi _{f}
\rangle \langle \Psi _{f}\mid \mid \hat{O}_{2}\mid \mid \Psi
_{0}\rangle \delta (E_{f}-E_{0}-\omega )} \nonumber
\end{eqnarray}
are the response functions with respect to the transition operators $\hat{O}
_{1}$ and $\hat{O}_{2}$ (when $\hat{O}_{1}=\hat{O}_{2}$ we use the notation $
R_{\hat{O}}=R_{\hat{O}\hat{O}}$). $\mid\Psi _{0,f}\rangle $ and $E_{0,f}$
are the wave function and energy of the ground and final state,
respectively. The transition operators $C_{J}(q), L_{J}(q),
E_{J}(q), M_{J}(q)$ are the reduced Coulomb, longitudinal, transverse
electric and transverse magnetic multipole operators. Since the
relevant energy regime is up to $\approx 60$ MeV, the main operators
contributing to the inelastic cross-section are the axial vector operators
$E^5_2, L^5_2, M^5_1, L^5_0 $ and the vector $C_1, E_1, L_1$. Usually, the main
contribution comes from the Gamow-Teller $E^5_1$ operator but due to the
closed shell character of the $^{4}\mathrm{He}$ nucleus, it is highly
suppressed. In this energy range the long wavelength
limit \cite{DO76} is accurate up to about $5\%$, thus it is highly
informative to look at the the long wavelength expansion of these
operators 
\begin{eqnarray}
{C_{1M}(q)=F_V \frac{qr}{3} Y_{1M}(\hat{r})} &&
\nonumber \\
{E_{1M}(q)=-\sqrt{2} \frac{\omega}{q}C_{1M}(q)} &&
\nonumber \\
{L_{1M}(q)=- \frac{\omega}{q}C_{1M}(q)} &&
\nonumber \\
{M_{1M}^5(q)=F_A \frac{qr}{3} \vec{\sigma} \cdot \vec{Y}_{11M}(\hat{r})} &&
\label{eq:lw} \\
{E_{2M}^5(q)=-i \sqrt{\frac{3}{5}} F_A \frac{qr}{3} \vec{\sigma} \cdot
\vec{Y}_{21M}(\hat{r})} &&
\nonumber \\
{L_{2M}^5(q)=\sqrt{\frac{2}{3}} E_{2M}^5(q)} &&
\nonumber \\
{L_{00}^5(q)=-i F_A \frac{qr}{3} \vec{\sigma} \cdot
\vec{Y}_{010}(\hat{r})}
\nonumber
\end{eqnarray}

\begin{figure}
\rotatebox{270}{
\resizebox{7cm}{!}{
\includegraphics{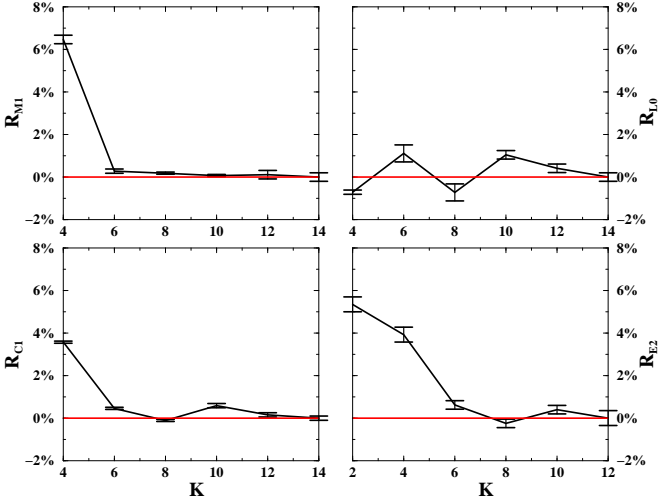} } }
\caption{\label{fig:conv} relative error in the sum-rule of the
leading response functions with respect to the hyper-angular momentum
quantum number $K$. The error bars reflect the uncertainty in inverting
the LIT.}
\end{figure}

Here $\vec{r}$ is the nucleon's location relative
to the system's center of mass.
The response functions are calculated by inverting the Lorentz
integral transforms 
\[
{L_{\hat{O}_{1}\hat{O}_{2}}(\sigma)=
\int d\omega \frac{R_{\hat{O}_{1}\hat{O}_{2}}(\omega )}{(\omega
-\sigma _{R})^{2}+\sigma _{I}^{2}}=\langle \tilde{\Psi}_{1}\mid \tilde{\Psi}
_{2}\rangle }, 
\]
where $\sigma =\sigma _{R}+i\sigma _{I}$, and $\mid\tilde{\Psi}_{i}\rangle $
($i=1,2$) are solutions of the Schr\"{o}dinger like equations 
\[
{(H-E_{0}-\sigma )\mid \tilde{\Psi}_{i}(\sigma )\rangle =\hat{O_{i}}\mid
\Psi _{0}\rangle }. 
\]
The localized character of the ground state, and the imaginary part of $
\sigma $, give these equations an asymptotic boundary condition
similar to a bound state. As a result, one can solve these equations
using the EIHH \cite{BA00} method. In this approach, the potential is
replaced by an effective potential constructed via the Lee-Suzuky
method \cite{LS80}, and the new equation is solved by expanding $\Psi
_{0}$ and $\tilde{\Psi}_{i}$ in four-body anti-symmetrized
Hyperspherical Harmonics basis functions \cite{BA97}. We calculate the
matrix element $\langle \tilde {\Psi}_{1}\mid \tilde{\Psi}_{2}\rangle$
using the Lanczos algorithm \cite{MA02}. 

The combination of the EIHH and LIT methods brings to a rapid
convergence in the Response functions. In Fig. ~\ref{fig:conv} , one
can see the relative error in the sum-rule of the main response
functions with respect to the hyper-angular momentum quantum number
$K$. It can be seen that upon convergence the relative error is well
below $1\%$. The error bars presented reflect the error in inverting
the LIT. Bearing in mind that the cross-section, up to kinematical
factors, is the sum of the response functions, this is a measure of the
accuracy in the calculation of the cross-section.
\begin{figure} [t]
\rotatebox{270}{
\resizebox{7.1cm}{!}{
\includegraphics{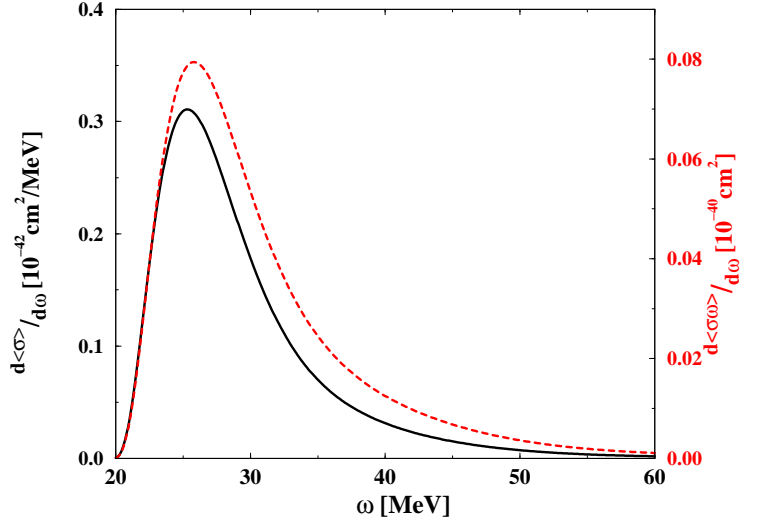}  } }
\caption{\label{fig:crs} Temperature averaged inelastic cross-sections at  
temperature $T=10$ MeV. The solid line is the differential cross-section, 
$\langle \frac{d\sigma}{d\omega}\rangle_T=\frac{1}{2}
\frac{1}{A} \langle \frac{d\sigma_\nu}{d\omega}
                   +\frac{d\sigma_{\overline{\nu}}}{d\omega}\rangle_T$,
(left scale). The dashed line is the differential
energy transfer cross-section,
$\langle \omega \frac{d\sigma}{d\omega}\rangle_T=\frac{1}{2}
\frac{1}{A} \langle \omega\frac{d\sigma_\nu}{d\omega}
                   +\omega\frac{d\sigma_{\overline{\nu}}}{d\omega}\rangle_T$,
 (right scale).
}
\end{figure}

It is well known that realistic 2--body NN potentials lead to an under-binding
of about $0.5-1$ MeV for the $^3$He and the triton nuclei and an 
under-binding of about $3-4$ MeV for $^4$He. For the AV8' force with a simple 
Coulomb interaction
we obtained a binding energy of $25.19$ MeV for $^4$He, and $7.76$ MeV for
the triton. Thus our model has a discrepancy, $\Delta \approx 2.4$ MeV, with 
respect to the experimental inelastic reaction threshold. In order to correct
for this difference we shifted the response function to the true threshold,
i.e. $R(\omega)\longrightarrow R(\omega-\Delta)$.

It is assumed that the neutrinos are in thermal equilibrium, thus
their spectrum can be approximated by the Fermi-Dirac distribution
with characteristic temperature $T$. As a result, the
interesting quantities are the temperature averaged cross-section and
energy transfer cross-section: 
\begin{eqnarray} 
{\frac{d\langle\sigma\rangle_T}{d\omega}=\int dk_i
f(T,k_i)\frac{d\sigma}{dk_f}} \label{eq:crs1} \\
{\frac{d\langle\sigma \omega \rangle_T}{d\omega}=\omega
\frac{d\langle\sigma\rangle_T}{d\omega}} \label{eq:rr}
\end{eqnarray} 
where $f(T,k)$ is normalized Fermi-Dirac spectrum with zero chemical
potential, temperature $T$, and energy $k$, i.e.,
\begin{eqnarray} 
{f(T,k)=\frac{0.5546}{T^3}\frac{k^2}{e^{k/T}+1}}.
\end{eqnarray} 
As a typical example we present in Fig.~\ref{fig:crs} the calculated
cross-section for $T=10$ MeV. 
In Table~\ref{tab:crs} we present the calculated total temperature
averaged cross-section, $\langle\sigma\rangle_T=\frac{1}{2}
\frac{1}{A} \langle \sigma_\nu+\sigma_{\overline{\nu}}\rangle_T$, and
energy transfer cross-section, $\langle\sigma \omega \rangle_T=\frac{1}{2}
\frac{1}{A} \langle \omega \sigma_\nu+ \omega \sigma_{\overline{\nu}}
\rangle_T$, as a function of the neutrinos' temperature. Also
presented are earlier results by Woosley et. al. \cite{WO90}. It can
be seen that the current work predicts an enhancement of about
$10\%-30\%$ in the cross-section. 

The energy transfer cross-section was fitted by Haxton to the formula
\cite{HA88}, 
\begin{equation}
\langle\sigma \omega \rangle_T = \alpha \left( \frac{T-T_0}{10 \rm{MeV}}\right)^\beta
\end{equation}
with the parameters $\alpha=0.62 \cdot \rm{10^{-40} cm^2 MeV}$,
$T_0=2.54 \rm {MeV}$, $\beta=3.82$. A similar fit to our results
yields $\alpha=0.64 \cdot \rm{10^{-40} cm^2 MeV}$, $T_0= 2.05 \rm
{MeV}$, $\beta=4.46$. It can be seen that the current work predicts a stronger
temperature dependence of the cross sections. For example, a $15\%$ differnce 
between
these calculations at $T=10$ Mev, grows to a $50\%$ difference at $T=16$ MeV.  

In conclusion, a detailed realistic calculation of the inelastic
neutrino-$^{4}\mathrm{He}$ neutral scattering cross-section is given. The
calculation was done in the impulse approximation with numerical
accuracy of about $1\%$. The different approximations used here should
result in about $10\%$ error, mainly due to many body currents, which
are not considered in the current work.

The effect of these results on the supernova explosion mechanism
should be checked through hydrodynamic simulations, of various
progenitors. Nonetheless, it is
clear that our results facilitate a stronger neutrino-matter coupling
in the supernova environment. First, our calculations predict
an enhanced cross section by $10\%-30\%$ with respect to previous estimates.
Second, we obtained steeper dependence of the energy transfer cross-section
on the neutrino's temperature. Thus, supporting the observation
that the core temperature is a critical parameter in the
explosion process. 
It is
important to notice that the energy-transfer due to inelastic
reactions are $1-2$ orders of magnitude larger than the elastic
reactions, ergo the inelastic cross-section are important to an
accurate description of the Helium shell temperature.

\begin{acknowledgments} 
The Authors would like thank S. Balberg, W. Leidemann, L. Marcucci, 
and E. Livne for their help and advice. 
This work was supported by the ISRAEL SCIENCE
FOUNDATION (grant no 202/02).
\end{acknowledgments}

\bibliography{nu_alpha_v2}
\end{document}